\documentclass[prb,aps,twocolumn,showpacs,amsmath,amssymb,floatfix,superscriptaddress,citeautoscript]{revtex4-1}
\usepackage{graphicx}
\usepackage{bm}
\usepackage{times}

\newcommand{\br}{{\bf r}}

\newcommand{\bi}{\mathbf}

\newcommand{\Eth}{E_{\text{Th}}}
\newcommand{\Volume}{{\cal A}}

\newcommand{\vf}{v_{\scriptscriptstyle F}}

\begin{document}

\title{Fermi-liquid regime of the mesoscopic Kondo problem}
\author{Denis Ullmo}
\email{denis.ullmo@u-psud.fr}
\affiliation{Univ Paris-Sud, LPTMS, UMR8626, 91405 Orsay, France, EU}
\affiliation{CNRS, 91405 Orsay, France, EU}
\author{Dong E. Liu}
\affiliation{Duke University, Box 90305,  Durham, NC 27708-0305, USA} 
\author{S\'{e}bastien Burdin}
\affiliation{Univ Bordeaux, LOMA, UMR 5798, F-33400 Talence, France, EU}
\affiliation{CNRS, LOMA, UMR 5798, F-33400 Talence, France, EU}
\author{Harold U. Baranger} 
\affiliation{Duke University, Box 90305,  Durham, NC 27708-0305, USA} 
\date{\today}

\begin{abstract}
  We consider the low temperature regime of the mesoscopic Kondo
  problem, and in particular the relevance of a Fermi-liquid
  description of this regime.  Using two complementary approaches -- a
  mean field slave fermion approximation on the one hand and a
  Fermi-liquid description ``\`a la Nozi\`eres'' supplemented by an
  argument of separation of scale on the other hand -- we show that
  they both lead to (essentially) the same quasi-particle spectra,
  providing in this way a strong indication that they both give
  the correct physics of this regime.
\end{abstract}

%71.10.Ay 	Fermi-liquid theory and other phenomenological models
%73.21.La 	Quantum dots 
%73.20.Hb 	Impurity and defect levels; energy states of adsorbed species 
%05.30.Fk 	Fermion systems and electron gas (see also 71.10.-w
%     Theories and models of many-electron systems; see also 67.10.Db
%     Fermion degeneracy in quantum fluids) 
%05.40.-a 	Fluctuation phenomena, random processes, noise, and
%     Brownian motion (for fluctuations in superconductivity, see 74.40.-n;
%     for statistical theory and fluctuations in nuclear reactions, see
%     24.60.-k; for fluctuations in plasma, see 52.25.Gj; for nonlinear
%     dynamics and chaos, see 05.45.-a) 
%05.45.Ac 	Low-dimensional chaos 
%
%75.20.Hr 	Local moment in compounds and alloys; Kondo effect,
%     valence fluctuations, heavy fermions (for Kondo effect and scattering
%     mechanisms in electronic conduction, see 72.15.Qm and 72.10.Fk) 
%
%
 
\maketitle

The concept of a Fermi liquid \cite{Fermiliquid1,PinesNozieresVol1},
introduced by Landau to describe the low temperature properties of
He$^3$ (above the superfluid transition), turned out to be one of the
most effective of condensed matter physics.  In its original
phenomenological version, its formulation is based on the assumption
that, in spite of possibly rather strong bare interactions, a set of
interacting fermions may (under certain conditions) behave at low
energy as weakly interacting quasi-particles.  For Landau
Fermi-liquids, such as the ones describing He$^3$ or the electron
liquid for $d>1$, a modern formulation in terms of renormalization
group analysis \cite{ShankarRMP94} provides a rigorous basis for this
hypothesis.  The main strength of the Fermi-liquid approach, however,
is that it makes it possible to take full advantage of the symmetries
of the problem under consideration, and of the fact that the
temperatures or energies considered are much smaller than all
``natural'' energy scales of the problem, allowing for perturbative
expansions near the Fermi energy.  In this way, the quasi-particles as
well as their weak mutual interactions can be characterized by a small
number of parameters, quite often fixed in practice by measuring a few
relevant quantities.  From those, the full behavior of the system can
be determined.

Another kind of Fermi liquid, which is going to be our main concern in
this paper, is the one introduced by Nozi\`eres
\cite{Fermiliquid1,Nozieres74,Nozieres80} to describe 
the low energy physics of the Kondo problem \cite{HewsonBook}.  In its simplest
version, referred to as the $s$-$d$ (or simply Kondo) model, this
consists in the (local) interaction of a gas of non-interacting fermions
with a spin one half. The corresponding Hamiltonian then reads
\begin{equation}
\label{eq:KondoHam}
H_{\rm K} = \sum_{\alpha \sigma} \epsilon_\alpha 
\hat c^{\dagger}_{\alpha \sigma} \hat c_{\alpha\sigma}
+ H_{\rm int} + g \mu_B B \cdot  S_z \; ,
\end{equation}
where $\hat c^{\dagger}_{\alpha \sigma}$ creates a particle with energy
$\epsilon_\alpha$, spin $\sigma$ and wave-function $\varphi_{\alpha}({\bf
  r})$, and the interaction with the impurity is expressed as 
\begin{equation}
\label{eq:KondoHam2}
H_{\rm int} = J_0 \, \bi{ S} \cdot \bi{s} (\br_0) 
\end{equation}
with $J_0 > 0$ the coupling strength, $ \hbar \mathbf{S} =
\hbar (S_x,S_y,S_z)$ a quantum spin $1/2$ operator ($S_i$ is half of the Pauli
matrix $\sigma_i$), $\hbar \mathbf{s}(\br_0) = ({\hbar}/{2}) \hat
\Psi^{\dagger}_{\sigma}(\br_0)\boldsymbol{\sigma}_{\sigma\sigma^\prime} \hat
\Psi_{\sigma}(\br_0)$ the spin density of the electron gas at the impurity
position ${\br_0}$, and $\hat \Psi^{\dagger}_{\sigma}(\br_0)=
\sum_{\alpha} \varphi_{\alpha}(\br_0) \hat
c^\dagger_{\alpha}$. Finally, with the last term on the right hand
side of Eq.\,(\ref{eq:KondoHam}) we consider also the possibility that
the impurity spin  is coupled to a magnetic field
$\mathbf{B} = B \mathbf{\hat z}$, with $g$ and $\mu_B$ the
corresponding Land\'e factor and Bohr magneton.

In the original formulation of the Kondo problem, the magnetic
impurity was assumed to be an actual impurity (e.g.\ Fe) in a bulk
piece of metal (e.g.\ Au), in which case the wave-functions
$\varphi_{\alpha}$ are plane waves and the spacing $\Delta$ between
the energy levels $\epsilon_\alpha$ can be assumed negligibly small
and constant.  The electron gas is then characterized by only two
quantities: the local density of states $\nu_0 = (\Volume
\Delta)^{-1}$ ($\Volume$ is the volume of the sample), and the
bandwidth $D_0$ of the spectrum.  The Kondo problem assumes the
dimensionless parameter $\mathcal{J}_0 \equiv \nu_0 J_0$ to be small.
In that case, it can be shown \cite{HewsonBook} that, beyond the Fermi energy
which essentially defines the origin of the energies, the problem is
characterized by a single energy scale, the Kondo temperature $T_K$
[which in a one-loop perturbative approximation is given by $T^\text{
  1-loop}_K = D_0 \exp(-1/(J_0\nu_0))$].  The Kondo temperature $T_K$
specifies the crossover between two different regimes: a weak
interaction regime for $T \gg T_K$ where, despite a
renormalization of the coupling constant, the magnetic impurity is
largely decoupled from the electron gas, and a strong interaction
regime $T \ll T_K$ where the magnetic impurity forms a singlet with
the electrons of the gas.

Nozi\`eres \cite{Nozieres74}, using some physical arguments related to the
large $\mathcal{J}_0$ limit of the Hamiltonian (\ref{eq:KondoHam}) and
evidence from numerical renormalization group calculations of
Wilson \cite{WilsonRMP75}, proposed that the low temperature regime of the Kondo
problem should be a Fermi liquid.  As for Landau Fermi-liquids,
symmetries, and the fact that the only energy scale is the Kondo
temperature, makes it possible to essentially completely specify the
properties of this Fermi liquid: the quasi-particles are
characterized by a phase shift $\delta_s(\epsilon_F) = s \pi/2$ at the
Fermi energy ($s=\pm 1$ is the sign of the spin), with a variation
\begin{equation} \label{eq:phase_shift}
\delta_s(\epsilon_F+\omega,B) = s \pi/2 + \omega /T_K - s (g \mu_B/2) B/T_K \; ,
\end{equation}
away from the Fermi energy and with a small magnetic field $B$.  In
addition to this phase shift, the magnetic impurity generates a weak
effective interaction between the electrons of the gas associated with
virtual breaking of the Kondo singlet, and thus taking place only locally at
the impurity, $V_{\rm eff} = (\pi\nu_0^2 T_K)^{-1} \hat
\Psi^{\dagger}_\uparrow(\br_0) \hat \Psi_\uparrow(\br_0) \hat
\Psi^{\dagger}_\downarrow(\br_0) \hat \Psi_\downarrow(\br_0)$.

The progress made in the control and miniaturization of micro- or
nano-structures has renewed interest in the Kondo problem, and in
particular made relevant the question of what would happen if the
magnetic impurity is connected to a {\em finite-size} fully-coherent
electron bath rather than a bulk piece of material
\cite{
Thimm99, Cornaglia02a,*Cornaglia02b,*Cornaglia03, Kang00,Affleck01, 
Simon02,*Simon03,KaulEPL05,Yoo05,KaulPRL06,Simon06,Pereira08,RotterAlhassid08,
RotterAlhassid09, Kaul09, Bedrich10, LiuEPL2012,LiuPRB2012}.  
As a first consequence, the finite size of the bath implies  a
finite mean level spacing $\Delta$, and it becomes meaningful to
discuss the individual properties of levels and wave-functions.  Lack
of translational invariance will furthermore be associated with
interference effects, and thus mesoscopic fluctuations
of the energy levels and wave-functions. These mesoscopic fluctuations
will affect the Kondo physics, and in particular imply
fluctuations of the Kondo temperatures \cite{Zarand96,Kettemann04,KaulEPL05,Yoo05, Kettemann06,Kettemann07,LiuPRB2012}.

The question we address in this paper is how the low
temperature Fermi-liquid description of the Kondo problem is modified
in the presence of mesoscopic fluctuations.  To simplify the
discussion, we consider the (rather typical) situation where
mesoscopic fluctuations exist for energies between the mean
level spacing $\Delta$ and some characteristic energy scale 
$E_{\rm fl}$ (such as the Thouless energy), and furthermore assume 
this latter energy scale to be smaller
than the Kondo temperature.  Thus as one lowers the
temperature, the Kondo singlet is formed before any mesoscopic
fluctuations set in, so that one does not expect any fluctuations of
the Kondo temperature itself (the interrelation between the
fluctuations of the Kondo temperature and the mesoscopic fluctuations
of the quasi-particles is left for future studies).

As always when one tries to develop a Fermi-liquid description for a
mesoscopic system, a conceptual difficulty arises.  As mentioned
above, what makes a Fermi liquid description effective is not so
much the derivation of its parameters from a microscopic
calculation, but rather the fact that because there are only a small number
of energy scales at play (and usually a high degree of symmetry), the
number of parameters required to fully describe the system is rather
small.  For mesoscopic systems, however, one has fluctuations at all
scales between $\Delta$ and $E_{\rm fl}$.  It is therefore a priori
hopeless to obtain a parameterization with only a small number of
parameters.

In such  circumstances, one possible approach to the
Fermi-liquid regime of the mesoscopic Kondo problem is to use a mean
field slave fermion approach. Up to some well understood limitations,
such a mean field approximation is known to provide a good description
of the Fermi-liquid regime of the bulk Kondo problem
\cite{HewsonBook}, and it is reasonable to expect that this will
remain the case for its  mesoscopic version.

Let us summarize briefly the principle of the mean field approximation
scheme\cite{MeanFieldForSpinReview}.  Starting from the Hamiltonian $H_K$
Eqs.\,(\ref{eq:KondoHam})-(\ref{eq:KondoHam2}), one introduces a
representation of the spin ${\bf S}$ in terms of Abrikosov fermions
$\hat f^\dagger_{\sigma}$ with spin $\sigma=\uparrow,\downarrow$ and writes
the Kondo part of the Hamiltonian as
\begin{equation}
H_{\rm int}=
\frac{J_{0}}{2}\sum_{\sigma\sigma'}
f_{\sigma}^{\dagger}f_{\sigma'} \hat \Psi^{\dagger}_{\sigma'}(\br_0)
\hat \Psi_{\sigma}(\br_0)
-
\frac{J_{0}}{4}\sum_{\sigma} \hat \Psi^{\dagger}_{\sigma}(\br_0) \hat
\Psi_{\sigma}(\br_0) .  
\end{equation}
One furthermore needs to impose  the constraint 
\begin{equation}
\hat f_{\uparrow}^{\dagger} \hat f_{\uparrow}
+ \hat f_{\downarrow}^{\dagger} \hat f_{\downarrow} = 1 \; ,
\label{eq_constraint}
\end{equation}
which is done
through 
%the introduction of 
a Lagrange multiplier $\epsilon_0$ in the
action and amounts in practice to the substitution
\[
H_{\rm imp}\mapsto H_{\rm imp}
+ \epsilon_0 \sum_{\sigma} \hat f_{\sigma}^{\dagger} \hat f_{\sigma} \; . 
\] 

This fermionic representation is exact.  The mean field approximation
consists in replacing the quartic part of the Kondo term by an
effective quadratic term $ \sum_{\sigma\sigma'}
\hat f_{\sigma}^{\dagger} \hat f_{\sigma'} \hat
\Psi^{\dagger}_{\sigma}(\br_0) \hat \Psi_{\sigma}(\br_0)
\mapsto \sum_{\sigma\sigma'} [\hat f_{\sigma}^{\dagger} \hat \Psi_{\sigma}(\br_0)
  \langle \hat \Psi^{\dagger}_{\sigma'}(\br_0) \hat f_{\sigma'}\rangle $ $+\;
  \text{h.c.}
%    \hat \Psi^{\dagger}_{\sigma}(\br_0) \hat f_{\sigma} \langle
%    \hat f_{\sigma'}^{\dagger} \hat \Psi_{\sigma'}(\br_0) \rangle 
 \big]$ where the mean
value $\langle\cdots\rangle$ is computed self consistently.
The mean field Hamiltonian obtained in this way is a resonant
level model:
\begin{eqnarray} 
 H_{\rm MF} & = & \sum_{\sigma} \Bigg[ 
\left( \sum_{\alpha} \epsilon_\alpha 
\hat c^{\dagger}_{\alpha \sigma} \hat c_{\alpha\sigma} \right) 
+ \epsilon_{0\sigma} \hat f_\sigma^\dagger \hat
f_\sigma
\nonumber  \\
&& 
+ \left( v^*  \hat f_\sigma^\dagger \hat \Psi_{\sigma}(\br_0)
+ v     \hat \Psi^{\dagger}_{\sigma}(\br_0)  \hat f_\sigma \right)
\Bigg] \label{eq:meanfield} 
\end{eqnarray}
where we define
%\begin{equation}
%\epsilon_{0\sigma}  = \epsilon_o + \sigma \frac{g \mu_B}{2} B.
%\end{equation}
$\epsilon_{0\sigma} \equiv \epsilon_o + s g \mu_B B/2 $.
To fix the parameters of the resonant level model, there are two
self-consistency conditions,  
\begin{eqnarray} 
v &  = & \frac{J_0}{2} \sum_{\sigma}
  \langle  \hat f_\sigma^\dagger \hat \Psi_{\sigma}(\br_0)  \rangle 
\; , \\
1 & = & \sum_\sigma\langle f_\sigma^\dagger \hat f_\sigma  \rangle \; ;
\end{eqnarray} 
note that the second one consists in treating the constraint
Eq.\,(\ref{eq_constraint}) only on average. This mean field treatment
has been used to study the mesoscopic Kondo problem in
Refs.\,\onlinecite{Bedrich10, LiuEPL2012,LiuPRB2012}. 

Under our assumption that the Kondo temperature is larger than the
scale $E_{\rm fl}$ below which mesoscopic fluctuations occur, we
expect that the parameters $v$ and $\epsilon_0$ are the same as for the
bulk analogue of our system.  A detailed analysis of the fluctuations
of the mean field parameters shows indeed that their relative variance
goes to zero as $T_K$ becomes much larger than the mean level spacing
$\Delta$ \cite{KaulEPL05,Kettemann07,LiuPRB2012}. Therefore, in the low temperature
regime, $\epsilon_0$ is fixed to the Fermi energy
$\mu$ (that we take equal to zero), and the low temperature
limit 
%$\Gamma(0)$ 
of the resonance width $\Gamma(T) \equiv \pi \nu_0
|v|^2 $ can be interpreted as the Kondo temperature.  More precisely,
$\Gamma(0) = a_k T_K^{\rm MF} = \frac{1}{2} T^\text{1-loop}_K \quad
(a_k \simeq 1.133\cdots)$, where $T_K^{\rm MF}$ and $T^\text{1-loop}_K$ are
respectively the mean field and one-loop approximations to the Kondo
temperature.  Now as can be easily shown (see
e.g.\ Ref.\,\onlinecite{LiuPRB2012}) the eigenenergies $\lambda_{\beta \sigma}$ of the resonant level are the solutions of the equation 
\begin{equation} \label{eq:specMF}
\sum_\alpha \frac{|\varphi_\alpha(\br_0)|^2}{\lambda -
  \epsilon_\alpha} = \frac{\lambda - \epsilon_{0\sigma}}{|v|^2} 
  = \frac{\pi \nu_0}{\Gamma} \left(\lambda - \epsilon_{0\sigma} \right) \; .
\end{equation}
In Refs.\ \onlinecite{LiuEPL2012,LiuPRB2012}, this equation was the
starting point for analyzing the spectral fluctuations of the
Landau quasi-particle energies of the low temperature mesoscopic Kondo problem.

What we intend to show here is that Eq.\,(\ref{eq:specMF}), or at least
its close analog, can be derived directly from a
Fermi-liquid analysis ``\`a la Nozi\`eres'' of the mesoscopic Kondo problem.
The approach we follow is similar in
spirit to the way a Landau Fermi-liquid description is obtained for a mesoscopic electron liquid, relevant in the context of
ballistic or diffusive quantum dots \cite{Aleiner02PhysRep}.  
Indeed in that case, too, the
existence of fluctuations at all energy scales between the Thouless
energy and the mean level spacing seems 
%a priori HUB: to include this, needs to be in italics because it is latin which then attracts too much attention to it, so I cut it...
to prevent the
description of the problem in terms of a few parameters.  This
problem has been ``solved'' using essentially an argument
of separation of scales: if the screening length
of the Coulomb interaction is much smaller than the size of
the dot, the renormalization of the mass and of the interactions (short
length scale) are the same as in the bulk, while the confinement effects
(large length scale) are obtained from a self-consistent potential
derived from a Thomas-Fermi treatment.  Note however that this
common wisdom way of dealing with a mesoscopic Landau Fermi liquid
does not rely on an analytic derivation (see for instance the
renormalization group approach proposed in
Ref.\,\onlinecite{Aleiner02PhysRep} and the difficulty with that
approach discussed in Ref.\,\onlinecite{UllmoRPP2008}), but it has 
been effective in interpreting most experimental data.

This is the philosophy we would like to follow for the Fermi-liquid
description of the mesoscopic Kondo problem.  To remain general, we
assume only that, as above, our mesoscopic system is characterized by
the energy scale $E_{\rm fl}$, below which mesoscopic fluctuations take
place but above which our electron bath behaves essentially as a bulk
system.  Thus, smoothing the density of states $\rho_B(\epsilon)
\equiv \sum_\alpha\delta(\epsilon - \epsilon_\alpha) $ on the scale
$E_{\rm fl}$ gives a ``bulk''-like density of states $\rho_0(\epsilon)
\equiv \langle \rho_B \rangle_{E_{\rm fl}}$ which shows no mesoscopic
fluctuations and has only secular variations on classical scales
(e.g.\ the Fermi energy).  In the same way, we assume that a bulk-like
wave-function probability $\gamma_0(\epsilon) \equiv \langle
|\varphi_\alpha|^2 \rangle_{E_{\rm fl}}$ can be defined with variation
only on classical scales.  To help visualize the problem, one may
think of the mesoscopic electron bath as a billiard, thus
corresponding to the one particle Schr\"odinger Hamiltonian $H_0 = -
(\hbar^2/2m) \nabla^2$ inside some domain $\mathcal{D}$ of area
$\Volume$ and typical size $L$. In that case, $\rho_0(\epsilon)$ is
the Weyl mean density of states, $\gamma_0(\epsilon) = 1/\Volume$, and
$E_{\rm fl}$ is the Thouless energy $\Eth = \hbar/\tau_{\rm fl}$ with
$\tau_{\rm fl} \equiv L / \vf$ the time of flight across the system
($\vf$ is the Fermi velocity). We shall not, however, use any specific
properties of billiards in what follows and shall furthermore assume
that, unlike billiards, our system has a finite bandwidth $D$ (which
avoids unessential technical convergence problems and is in any case
necessary for a properly defined Kondo problem).

Under the assumption that the Kondo temperature is larger
than the Thouless energy, 
%one can consider that 
the impurity behaves at the local level as if it were placed in a bulk
piece of material 
characterized by the density of states $\rho_0$ and the wave-function
probability $\gamma_0(\epsilon)$; it is only at a longer length
scale that the effects of the finiteness of the system are felt.  To
implement this intuitive idea,
%into a definite prescription 
let us for a moment consider 
%what would happen for 
a localized static potential
$U(\br) = U_0 \delta(\br -\br_0)$ (with $\br_0$ the location of the
impurity).  [The divergences associated with a $\delta$-potential are
avoided with a finite bandwidth.]  Let us furthermore denote by
$G_B(\br,\br';\omega)$ the Green function of our mesoscopic electron
bath, and by
$G_0(\br,\br';\omega)$ the corresponding ``bulk-like'' Green
function.  In particular we have 
\begin{eqnarray}
G_B(\br_0,\br_0;\omega) & = & \sum_\alpha 
\frac{|\varphi_\alpha(\br_0)|^2}{\omega - \epsilon_\alpha + i\eta} \\
G_0(\br_0,\br_0;\omega) & = & -i \pi \nu_0(\omega)  +  \Lambda(\omega)
\; ,
\end{eqnarray}
with $\nu_0(\omega) \equiv \gamma_0(\omega) \rho_0(\omega)$ the local density
of states, and 
\begin{equation}
 \Lambda(\omega) = \gamma_0(\omega) \cdot \mathcal{P} \int d \epsilon
\frac{ \rho_0(\epsilon)}{\omega - \epsilon} \; .
\end{equation}

The $T$-matrix of the static impurity in the bulk-like system is then given
by
\begin{eqnarray}
t(\omega) & = & U + UG_0U + UG_0UG_0U + \cdots \nonumber \\
          & = & U\frac {1}{1 -  G_0U} \; ,
\end{eqnarray} 
while for the genuine mesoscopic system one has similarly
 \begin{eqnarray}
T(\omega) & = & U + UG_BU + UG_BUG_BU + \cdots \nonumber \\
          & = & U\frac {1}{1 -  G_BU} \; .
\end{eqnarray}
Since $U_0$, $t(\omega)$, and $T(\omega)$ as well as
$G_0(\br_0,\br_0)$ and $G_B(\br_0,\br_0)$ are just numbers, simple
algebra leads to 
\begin{equation} \label{eq:Tmatrix}
T(\omega) = \frac {t(\omega)}{1 -  \delta G(\omega) t(\omega)} \; ,
\end{equation}
with  $\delta G \equiv G_B - G_0$ the fluctuating part of the 
Green function. From this equation the full Green function $G^{\rm tot}_B$ including both the
confinement  and the impurity potential is given by 
\begin{eqnarray} \label{eq:Gmatrix}
G^{\rm tot}_B(\br,\br';\omega) & = & G_B(\br,\br';\omega) \nonumber\\
& + & G_B(\br,\br_0;\omega) T(\omega) G_B(\br_0,\br';\omega) \; .
\end{eqnarray}

We argue that Eqs.~(\ref{eq:Tmatrix}) and
(\ref{eq:Gmatrix}) can be used 
%as a basis 
to characterize the
quasi-particles in the Fermi liquid description of the mesoscopic
Kondo problem.  Indeed, here the Green function $G_B$ contains all the
required information about the confinement properties (``long
range''), and the Kondo physics (``short range'') is implemented by
the $T$-matrix $t(\omega)$. At energies $\omega$ sufficiently small 
compared to $T_K$ so that inelastic processes are negligible, $t(\omega)$ is
related to the phase shift Eq.\,(\ref{eq:phase_shift}) through
\begin{equation} \label{eq:t(w)}
t(\omega) = -\frac{1}{2i \pi \nu_0} \big[\exp(2i\delta_s(\omega)) - 1
  \big] \; .
\end{equation}
The full Green function of the quasi-particles is therefore entirely
determined through Eq.~(\ref{eq:Gmatrix}).  In particular, the
quasi-particles energies $\lambda_\beta$ of the Fermi liquid are given
by the poles of the $T$-matrix (\ref{eq:Tmatrix}), and therefore fulfill
the equation
\begin{equation} \label{eq:spec1}
 G_B(\br_0,\br_0;\lambda) - G_0(\br_0,\br_0;\lambda) =
 1/t(\lambda) \; . 
\end{equation}
Noting that $\mathrm{Im} [1/t(\omega)] = \pi \nu_0$ and thus that away
from the energies $\epsilon_\alpha$ of the unperturbed problem 
the imaginary part of Eq.\,(\ref{eq:spec1}) is automatically fulfilled,
we see that the $\lambda$'s are therefore given as the solutions of
\begin{eqnarray} \label{eq:spec2} \sum_\alpha
  \frac{|\varphi_\alpha(\br_0)|^2}{\lambda - \epsilon_\alpha} -
  \Lambda(\lambda) & = & - \frac{\pi \nu_0 \sin(2\delta_s)}{1 -
    \cos(2\delta_s)} \nonumber\\
    & \simeq & \frac{\pi \nu_0}{T_K} 
\left(\lambda - s\frac{g\mu_B}{2}B\right) \; ,
\end{eqnarray}
where in the last equality we have inserted the expression
Eq.\,(\ref{eq:phase_shift}) for the phase shift and, to remain
consistent, expanded to first order in $1/T_k$.

Comparing Eqs.~(\ref{eq:specMF}) and (\ref{eq:spec2}), we see that they
have the same structure, and basically contain the same qualitative
content.  Quantitatively they differ in two respects.  First,
the true Kondo temperature $T_K$ in (\ref{eq:spec2}) is replaced in
Eq.\,(\ref{eq:specMF}) by the resonance width $\Gamma(0)$, i.e., up to
the factor $a_k \simeq 1.133\cdots$ by the mean field approximation
$T_K^{\rm MF}$ of the Kondo temperature.  This is of course expected
but should be kept in mind.  Second, the term $\Lambda(\lambda)$ in
(\ref{eq:spec2}) is absent from (\ref{eq:specMF}).  If the Fermi
energy is in the middle of the band, this is of little importance as
then $\Lambda \!\sim\! 0$.  If this is not the case, however, 
$\Lambda(\lambda)$
compensates the effect of states far from the Fermi energy, and its
absence in (\ref{eq:specMF}) is certainly a limitation of the mean
field approach.  We see therefore that both the mean field and
the Fermi liquid approaches have the same physical content as far as the
spectrum of the quasi-particles is concerned, but one
can expect better quantitative accuracy from the latter.

Finally, we can roughly estimate the range of energies for which the
above neglect of inelastic processes is applicable. It is known that
at low energy and temperature, the rate of inelastic processes in the
Kondo problem grows as the square of the deviation from the Fermi
energy. This is the expected Fermi liquid result; it has been shown,
for instance, that the imaginary part of the self-energy of the
quasi-particles goes to zero as $\omega^2$, $\Im \{\Sigma(\omega)\}
\propto \omega^2/T_K$.\cite{HewsonBook2} To extract the quasi-particle
energies via the procedure outlined here, this inelastic rate must be
much smaller than the level spacing, $\omega^2/T_K \!\ll\!
\Delta$. Thus, the number of quasi-particle energy levels that are
accurately treated by our argument, $N \!\equiv\! \omega/\Delta$, is
roughly given by $N \!\sim\!  \sqrt{T_K/\Delta}$. Since we are in any
case assuming $T_K \!\gg\!  \Delta$, this means that only levels near
the center of the Kondo resonance can be captured.

To conclude, we have shown that the Fermi-liquid regime of the
mesoscopic Kondo problem can be approached in two different ways:
within a slave-fermion--mean-field framework, or more directly from
a Fermi-liquid treatment ``\`a la Nozi\`eres''.  Limiting our discussion
here to the quasi-particle spectra, we have seen that if the chemical
potential is in the middle of the band, both approaches give the same
result, except that the true Kondo temperature is replaced in the
mean field approach by its mean field approximation (which is of
course the best the mean field can provide).  That both descriptions
are essentially equivalent is a strong indication that 
they provide the correct physics of this regime.  A
definitive confirmation of this statement could be obtained for
instance from a numerical renormalization group calculation for a few
mesoscopic realizations.

The work at Duke was supported by US DOE, Office of Basic Energy
Sciences, Division of Materials Sciences and Engineering under Grant
No.\,DE-SC0005237.
% (D.E.L. and H.U.B.).

%\bibliography{kondo,rmt,nano,general_ref,quantum-chaos}

\begin{thebibliography}{34}%
\makeatletter
\providecommand \@ifxundefined [1]{%
 \@ifx{#1\undefined}
}%
\providecommand \@ifnum [1]{%
 \ifnum #1\expandafter \@firstoftwo
 \else \expandafter \@secondoftwo
 \fi
}%
\providecommand \@ifx [1]{%
 \ifx #1\expandafter \@firstoftwo
 \else \expandafter \@secondoftwo
 \fi
}%
\providecommand \natexlab [1]{#1}%
\providecommand \enquote  [1]{``#1''}%
\providecommand \bibnamefont  [1]{#1}%
\providecommand \bibfnamefont [1]{#1}%
\providecommand \citenamefont [1]{#1}%
\providecommand \href@noop [0]{\@secondoftwo}%
\providecommand \href [0]{\begingroup \@sanitize@url \@href}%
\providecommand \@href[1]{\@@startlink{#1}\@@href}%
\providecommand \@@href[1]{\endgroup#1\@@endlink}%
\providecommand \@sanitize@url [0]{\catcode `\\12\catcode `\$12\catcode
  `\&12\catcode `\#12\catcode `\^12\catcode `\_12\catcode `\%12\relax}%
\providecommand \@@startlink[1]{}%
\providecommand \@@endlink[0]{}%
\providecommand \url  [0]{\begingroup\@sanitize@url \@url }%
\providecommand \@url [1]{\endgroup\@href {#1}{\urlprefix }}%
\providecommand \urlprefix  [0]{URL }%
\providecommand \Eprint [0]{\href }%
\providecommand \doibase [0]{http://dx.doi.org/}%
\providecommand \selectlanguage [0]{\@gobble}%
\providecommand \bibinfo  [0]{\@secondoftwo}%
\providecommand \bibfield  [0]{\@secondoftwo}%
\providecommand \translation [1]{[#1]}%
\providecommand \BibitemOpen [0]{}%
\providecommand \bibitemStop [0]{}%
\providecommand \bibitemNoStop [0]{.\EOS\space}%
\providecommand \EOS [0]{\spacefactor3000\relax}%
\providecommand \BibitemShut  [1]{\csname bibitem#1\endcsname}%
\let\auto@bib@innerbib\@empty
%</preamble>
\bibitem [{\citenamefont {Landau}(1957)}]{Fermiliquid1}%
  \BibitemOpen
  \bibfield  {author} {\bibinfo {author} {\bibfnamefont {L.~D.}\ \bibnamefont
  {Landau}},\ }\href@noop {} {\bibfield  {journal} {\bibinfo  {journal} {JETP}\
  }\textbf {\bibinfo {volume} {3}},\ \bibinfo {pages} {920} (\bibinfo {year}
  {1957})}\BibitemShut {NoStop}%
\bibitem [{\citenamefont {Pines}\ and\ \citenamefont
  {Nozi\`eres}(1966)}]{PinesNozieresVol1}%
  \BibitemOpen
  \bibfield  {author} {\bibinfo {author} {\bibfnamefont {D.}~\bibnamefont
  {Pines}}\ and\ \bibinfo {author} {\bibfnamefont {P.}~\bibnamefont
  {Nozi\`eres}},\ }\href@noop {} {\emph {\bibinfo {title} {Theory of Quantum
  Liquids Vol. I.}}}\ (\bibinfo  {publisher} {W. A. Benjamin},\ \bibinfo
  {address} {New York},\ \bibinfo {year} {1966})\BibitemShut {NoStop}%
\bibitem [{\citenamefont {Shankar}(1994)}]{ShankarRMP94}%
  \BibitemOpen
  \bibfield  {author} {\bibinfo {author} {\bibfnamefont {R.}~\bibnamefont
  {Shankar}},\ }\href@noop {} {\bibfield  {journal} {\bibinfo  {journal} {Rev.
  Mod. Phys.}\ }\textbf {\bibinfo {volume} {66}},\ \bibinfo {pages} {129}
  (\bibinfo {year} {1994})}\BibitemShut {NoStop}%
\bibitem [{\citenamefont {Nozi\`eres}(1974)}]{Nozieres74}%
  \BibitemOpen
  \bibfield  {author} {\bibinfo {author} {\bibfnamefont {P.}~\bibnamefont
  {Nozi\`eres}},\ }\href@noop {} {\bibfield  {journal} {\bibinfo  {journal} {J.
  Low Temp. Phys.}\ }\textbf {\bibinfo {volume} {17}},\ \bibinfo {pages} {31}
  (\bibinfo {year} {1974})}\BibitemShut {NoStop}%
\bibitem [{\citenamefont {Nozi\`eres}\ and\ \citenamefont
  {Blandin}(1980)}]{Nozieres80}%
  \BibitemOpen
  \bibfield  {author} {\bibinfo {author} {\bibfnamefont {P.}~\bibnamefont
  {Nozi\`eres}}\ and\ \bibinfo {author} {\bibfnamefont {A.}~\bibnamefont
  {Blandin}},\ }\href@noop {} {\bibfield  {journal} {\bibinfo  {journal} {J.
  Phys. France}\ }\textbf {\bibinfo {volume} {41}},\ \bibinfo {pages} {193}
  (\bibinfo {year} {1980})}\BibitemShut {NoStop}%
\bibitem [{\citenamefont {Hewson}(1993{\natexlab{a}})}]{HewsonBook}%
  \BibitemOpen
  \bibfield  {author} {\bibinfo {author} {\bibfnamefont {A.}~\bibnamefont
  {Hewson}},\ }\href@noop {} {\emph {\bibinfo {title} {The Kondo Problem to
  Heavy Fermions}}}\ (\bibinfo  {publisher} {Cambridge University Press},\
  \bibinfo {address} {Cambridge},\ \bibinfo {year} {1993})\BibitemShut
  {NoStop}%
\bibitem [{\citenamefont {Wilson}(1975)}]{WilsonRMP75}%
  \BibitemOpen
  \bibfield  {author} {\bibinfo {author} {\bibfnamefont {K.~G.}\ \bibnamefont
  {Wilson}},\ }\href {\doibase 10.1103/RevModPhys.47.773} {\bibfield  {journal}
  {\bibinfo  {journal} {Rev. Mod. Phys.}\ }\textbf {\bibinfo {volume} {47}},\
  \bibinfo {pages} {773} (\bibinfo {year} {1975})}\BibitemShut {NoStop}%
\bibitem [{\citenamefont {Thimm}\ \emph {et~al.}(1999)\citenamefont {Thimm},
  \citenamefont {Kroha},\ and\ \citenamefont {von Delft}}]{Thimm99}%
  \BibitemOpen
  \bibfield  {author} {\bibinfo {author} {\bibfnamefont {W.~B.}\ \bibnamefont
  {Thimm}}, \bibinfo {author} {\bibfnamefont {J.}~\bibnamefont {Kroha}}, \ and\
  \bibinfo {author} {\bibfnamefont {J.}~\bibnamefont {von Delft}},\ }\href
  {\doibase 10.1103/PhysRevLett.82.2143} {\bibfield  {journal} {\bibinfo
  {journal} {Phys. Rev. Lett.}\ }\textbf {\bibinfo {volume} {82}},\ \bibinfo
  {pages} {2143} (\bibinfo {year} {1999})}\BibitemShut {NoStop}%
\bibitem [{\citenamefont {Cornaglia}\ and\ \citenamefont
  {Balseiro}(2002{\natexlab{a}})}]{Cornaglia02a}%
  \BibitemOpen
  \bibfield  {author} {\bibinfo {author} {\bibfnamefont {P.~S.}\ \bibnamefont
  {Cornaglia}}\ and\ \bibinfo {author} {\bibfnamefont {C.~A.}\ \bibnamefont
  {Balseiro}},\ }\href@noop {} {\bibfield  {journal} {\bibinfo  {journal}
  {Phys. Rev. B}\ }\textbf {\bibinfo {volume} {66}},\ \bibinfo {pages} {115303}
  (\bibinfo {year} {2002}{\natexlab{a}})}\BibitemShut {NoStop}%
\bibitem [{\citenamefont {Cornaglia}\ and\ \citenamefont
  {Balseiro}(2002{\natexlab{b}})}]{Cornaglia02b}%
  \BibitemOpen
  \bibfield  {author} {\bibinfo {author} {\bibfnamefont {P.~S.}\ \bibnamefont
  {Cornaglia}}\ and\ \bibinfo {author} {\bibfnamefont {C.~A.}\ \bibnamefont
  {Balseiro}},\ }\href@noop {} {\bibfield  {journal} {\bibinfo  {journal}
  {Phys. Rev. B}\ }\textbf {\bibinfo {volume} {66}},\ \bibinfo {pages} {174404}
  (\bibinfo {year} {2002}{\natexlab{b}})}\BibitemShut {NoStop}%
\bibitem [{\citenamefont {Cornaglia}\ and\ \citenamefont
  {Balseiro}(2003)}]{Cornaglia03}%
  \BibitemOpen
  \bibfield  {author} {\bibinfo {author} {\bibfnamefont {P.~S.}\ \bibnamefont
  {Cornaglia}}\ and\ \bibinfo {author} {\bibfnamefont {C.~A.}\ \bibnamefont
  {Balseiro}},\ }\href@noop {} {\bibfield  {journal} {\bibinfo  {journal}
  {Phys. Rev. Lett.}\ }\textbf {\bibinfo {volume} {90}},\ \bibinfo {pages}
  {216801} (\bibinfo {year} {2003})}\BibitemShut {NoStop}%
\bibitem [{\citenamefont {Kang}\ and\ \citenamefont {Shin}(2000)}]{Kang00}%
  \BibitemOpen
  \bibfield  {author} {\bibinfo {author} {\bibfnamefont {K.}~\bibnamefont
  {Kang}}\ and\ \bibinfo {author} {\bibfnamefont {S.-C.}\ \bibnamefont
  {Shin}},\ }\href@noop {} {\bibfield  {journal} {\bibinfo  {journal} {Phys.
  Rev. Lett.}\ }\textbf {\bibinfo {volume} {85}},\ \bibinfo {pages} {5619}
  (\bibinfo {year} {2000})}\BibitemShut {NoStop}%
\bibitem [{\citenamefont {Affleck}\ and\ \citenamefont
  {Simon}(2001)}]{Affleck01}%
  \BibitemOpen
  \bibfield  {author} {\bibinfo {author} {\bibfnamefont {I.}~\bibnamefont
  {Affleck}}\ and\ \bibinfo {author} {\bibfnamefont {P.}~\bibnamefont
  {Simon}},\ }\href {\doibase 10.1103/PhysRevLett.86.2854} {\bibfield
  {journal} {\bibinfo  {journal} {Phys. Rev. Lett.}\ }\textbf {\bibinfo
  {volume} {86}},\ \bibinfo {pages} {2854} (\bibinfo {year}
  {2001})}\BibitemShut {NoStop}%
\bibitem [{\citenamefont {Simon}\ and\ \citenamefont
  {Affleck}(2002)}]{Simon02}%
  \BibitemOpen
  \bibfield  {author} {\bibinfo {author} {\bibfnamefont {P.}~\bibnamefont
  {Simon}}\ and\ \bibinfo {author} {\bibfnamefont {I.}~\bibnamefont
  {Affleck}},\ }\href {\doibase 10.1103/PhysRevLett.89.206602} {\bibfield
  {journal} {\bibinfo  {journal} {Phys. Rev. Lett.}\ }\textbf {\bibinfo
  {volume} {89}},\ \bibinfo {pages} {206602} (\bibinfo {year}
  {2002})}\BibitemShut {NoStop}%
\bibitem [{\citenamefont {Simon}\ and\ \citenamefont
  {Affleck}(2003)}]{Simon03}%
  \BibitemOpen
  \bibfield  {author} {\bibinfo {author} {\bibfnamefont {P.}~\bibnamefont
  {Simon}}\ and\ \bibinfo {author} {\bibfnamefont {I.}~\bibnamefont
  {Affleck}},\ }\href@noop {} {\bibfield  {journal} {\bibinfo  {journal} {Phys.
  Rev. B}\ }\textbf {\bibinfo {volume} {68}},\ \bibinfo {pages} {115304}
  (\bibinfo {year} {2003})}\BibitemShut {NoStop}%
\bibitem [{\citenamefont {Kaul}\ \emph {et~al.}(2005)\citenamefont {Kaul},
  \citenamefont {Ullmo}, \citenamefont {Chandrasekharan},\ and\ \citenamefont
  {Baranger}}]{KaulEPL05}%
  \BibitemOpen
  \bibfield  {author} {\bibinfo {author} {\bibfnamefont {R.~K.}\ \bibnamefont
  {Kaul}}, \bibinfo {author} {\bibfnamefont {D.}~\bibnamefont {Ullmo}},
  \bibinfo {author} {\bibfnamefont {S.}~\bibnamefont {Chandrasekharan}}, \ and\
  \bibinfo {author} {\bibfnamefont {H.~U.}\ \bibnamefont {Baranger}},\
  }\href@noop {} {\bibfield  {journal} {\bibinfo  {journal} {Europhys. Lett.}\
  }\textbf {\bibinfo {volume} {71}},\ \bibinfo {pages} {973} (\bibinfo {year}
  {2005})}\BibitemShut {NoStop}%
\bibitem [{\citenamefont {Yoo}\ \emph {et~al.}(2005)\citenamefont {Yoo},
  \citenamefont {Chandrasekharan}, \citenamefont {Kaul}, \citenamefont
  {Ullmo},\ and\ \citenamefont {Baranger}}]{Yoo05}%
  \BibitemOpen
  \bibfield  {author} {\bibinfo {author} {\bibfnamefont {J.}~\bibnamefont
  {Yoo}}, \bibinfo {author} {\bibfnamefont {S.}~\bibnamefont
  {Chandrasekharan}}, \bibinfo {author} {\bibfnamefont {R.~K.}\ \bibnamefont
  {Kaul}}, \bibinfo {author} {\bibfnamefont {D.}~\bibnamefont {Ullmo}}, \ and\
  \bibinfo {author} {\bibfnamefont {H.~U.}\ \bibnamefont {Baranger}},\
  }\href@noop {} {\bibfield  {journal} {\bibinfo  {journal} {Phys. Rev. B}\
  }\textbf {\bibinfo {volume} {71}},\ \bibinfo {pages} {201309(R)} (\bibinfo
  {year} {2005})}\BibitemShut {NoStop}%
\bibitem [{\citenamefont {Kaul}\ \emph {et~al.}(2006)\citenamefont {Kaul},
  \citenamefont {Zar\'and}, \citenamefont {Chandrasekharan}, \citenamefont
  {Ullmo},\ and\ \citenamefont {Baranger}}]{KaulPRL06}%
  \BibitemOpen
  \bibfield  {author} {\bibinfo {author} {\bibfnamefont {R.~K.}\ \bibnamefont
  {Kaul}}, \bibinfo {author} {\bibfnamefont {G.}~\bibnamefont {Zar\'and}},
  \bibinfo {author} {\bibfnamefont {S.}~\bibnamefont {Chandrasekharan}},
  \bibinfo {author} {\bibfnamefont {D.}~\bibnamefont {Ullmo}}, \ and\ \bibinfo
  {author} {\bibfnamefont {H.~U.}\ \bibnamefont {Baranger}},\ }\href@noop {}
  {\bibfield  {journal} {\bibinfo  {journal} {Phys. Rev. Lett.}\ }\textbf
  {\bibinfo {volume} {96}},\ \bibinfo {pages} {176802} (\bibinfo {year}
  {2006})}\BibitemShut {NoStop}%
\bibitem [{\citenamefont {Simon}\ \emph {et~al.}(2006)\citenamefont {Simon},
  \citenamefont {Salomez},\ and\ \citenamefont {Feinberg}}]{Simon06}%
  \BibitemOpen
  \bibfield  {author} {\bibinfo {author} {\bibfnamefont {P.}~\bibnamefont
  {Simon}}, \bibinfo {author} {\bibfnamefont {J.}~\bibnamefont {Salomez}}, \
  and\ \bibinfo {author} {\bibfnamefont {D.}~\bibnamefont {Feinberg}},\
  }\href@noop {} {\bibfield  {journal} {\bibinfo  {journal} {Phys. Rev. B}\
  }\textbf {\bibinfo {volume} {73}},\ \bibinfo {pages} {205325} (\bibinfo
  {year} {2006})}\BibitemShut {NoStop}%
\bibitem [{\citenamefont {Pereira}\ \emph {et~al.}(2008)\citenamefont
  {Pereira}, \citenamefont {Laflorencie}, \citenamefont {Affleck},\ and\
  \citenamefont {Halperin}}]{Pereira08}%
  \BibitemOpen
  \bibfield  {author} {\bibinfo {author} {\bibfnamefont {R.~G.}\ \bibnamefont
  {Pereira}}, \bibinfo {author} {\bibfnamefont {N.}~\bibnamefont
  {Laflorencie}}, \bibinfo {author} {\bibfnamefont {I.}~\bibnamefont
  {Affleck}}, \ and\ \bibinfo {author} {\bibfnamefont {B.~I.}\ \bibnamefont
  {Halperin}},\ }\href {\doibase 10.1103/PhysRevB.77.125327} {\bibfield
  {journal} {\bibinfo  {journal} {Phys. Rev. B}\ }\textbf {\bibinfo {volume}
  {77}},\ \bibinfo {pages} {125327} (\bibinfo {year} {2008})}\BibitemShut
  {NoStop}%
\bibitem [{\citenamefont {Rotter}\ \emph {et~al.}(2008)\citenamefont {Rotter},
  \citenamefont {T\"ureci}, \citenamefont {Alhassid},\ and\ \citenamefont
  {Stone}}]{RotterAlhassid08}%
  \BibitemOpen
  \bibfield  {author} {\bibinfo {author} {\bibfnamefont {S.}~\bibnamefont
  {Rotter}}, \bibinfo {author} {\bibfnamefont {H.~E.}\ \bibnamefont
  {T\"ureci}}, \bibinfo {author} {\bibfnamefont {Y.}~\bibnamefont {Alhassid}},
  \ and\ \bibinfo {author} {\bibfnamefont {A.~D.}\ \bibnamefont {Stone}},\
  }\href {\doibase 10.1103/PhysRevLett.100.166601} {\bibfield  {journal}
  {\bibinfo  {journal} {Phys. Rev. Lett.}\ }\textbf {\bibinfo {volume} {100}},\
  \bibinfo {pages} {166601} (\bibinfo {year} {2008})}\BibitemShut {NoStop}%
\bibitem [{\citenamefont {Rotter}\ and\ \citenamefont
  {Alhassid}(2009)}]{RotterAlhassid09}%
  \BibitemOpen
  \bibfield  {author} {\bibinfo {author} {\bibfnamefont {S.}~\bibnamefont
  {Rotter}}\ and\ \bibinfo {author} {\bibfnamefont {Y.}~\bibnamefont
  {Alhassid}},\ }\href {\doibase 10.1103/PhysRevB.80.184404} {\bibfield
  {journal} {\bibinfo  {journal} {Phys. Rev. B}\ }\textbf {\bibinfo {volume}
  {80}},\ \bibinfo {pages} {184404} (\bibinfo {year} {2009})}\BibitemShut
  {NoStop}%
\bibitem [{\citenamefont {Kaul}\ \emph {et~al.}(2009)\citenamefont {Kaul},
  \citenamefont {Ullmo}, \citenamefont {Zar\'and}, \citenamefont
  {Chandrasekharan},\ and\ \citenamefont {Baranger}}]{Kaul09}%
  \BibitemOpen
  \bibfield  {author} {\bibinfo {author} {\bibfnamefont {R.~K.}\ \bibnamefont
  {Kaul}}, \bibinfo {author} {\bibfnamefont {D.}~\bibnamefont {Ullmo}},
  \bibinfo {author} {\bibfnamefont {G.}~\bibnamefont {Zar\'and}}, \bibinfo
  {author} {\bibfnamefont {S.}~\bibnamefont {Chandrasekharan}}, \ and\ \bibinfo
  {author} {\bibfnamefont {H.~U.}\ \bibnamefont {Baranger}},\ }\href@noop {}
  {\bibfield  {journal} {\bibinfo  {journal} {Phys. Rev. B}\ }\textbf {\bibinfo
  {volume} {80}},\ \bibinfo {pages} {035318} (\bibinfo {year}
  {2009})}\BibitemShut {NoStop}%
\bibitem [{\citenamefont {Bedrich}\ \emph {et~al.}(2010)\citenamefont
  {Bedrich}, \citenamefont {Burdin},\ and\ \citenamefont
  {Hentschel}}]{Bedrich10}%
  \BibitemOpen
  \bibfield  {author} {\bibinfo {author} {\bibfnamefont {R.}~\bibnamefont
  {Bedrich}}, \bibinfo {author} {\bibfnamefont {S.}~\bibnamefont {Burdin}}, \
  and\ \bibinfo {author} {\bibfnamefont {M.}~\bibnamefont {Hentschel}},\ }\href
  {\doibase 10.1103/PhysRevB.81.174406} {\bibfield  {journal} {\bibinfo
  {journal} {Phys. Rev. B}\ }\textbf {\bibinfo {volume} {81}},\ \bibinfo
  {pages} {174406} (\bibinfo {year} {2010})}\BibitemShut {NoStop}%
\bibitem [{\citenamefont {Liu}\ \emph {et~al.}(2012{\natexlab{a}})\citenamefont
  {Liu}, \citenamefont {Burdin}, \citenamefont {Baranger},\ and\ \citenamefont
  {Ullmo}}]{LiuEPL2012}%
  \BibitemOpen
  \bibfield  {author} {\bibinfo {author} {\bibfnamefont {D.~E.}\ \bibnamefont
  {Liu}}, \bibinfo {author} {\bibfnamefont {S.}~\bibnamefont {Burdin}},
  \bibinfo {author} {\bibfnamefont {H.~U.}\ \bibnamefont {Baranger}}, \ and\
  \bibinfo {author} {\bibfnamefont {D.}~\bibnamefont {Ullmo}},\ }\href@noop {}
  {\bibfield  {journal} {\bibinfo  {journal} {Europhys. Lett.}\ }\textbf
  {\bibinfo {volume} {97}},\ \bibinfo {pages} {17006} (\bibinfo {year}
  {2012}{\natexlab{a}})}\BibitemShut {NoStop}%
\bibitem [{\citenamefont {Liu}\ \emph {et~al.}(2012{\natexlab{b}})\citenamefont
  {Liu}, \citenamefont {Burdin}, \citenamefont {Baranger},\ and\ \citenamefont
  {Ullmo}}]{LiuPRB2012}%
  \BibitemOpen
  \bibfield  {author} {\bibinfo {author} {\bibfnamefont {D.~E.}\ \bibnamefont
  {Liu}}, \bibinfo {author} {\bibfnamefont {S.}~\bibnamefont {Burdin}},
  \bibinfo {author} {\bibfnamefont {H.~U.}\ \bibnamefont {Baranger}}, \ and\
  \bibinfo {author} {\bibfnamefont {D.}~\bibnamefont {Ullmo}},\ }\href@noop {}
  {\bibfield  {journal} {\bibinfo  {journal} {Phys. Rev. B}\ }\textbf {\bibinfo
  {volume} {85}},\ \bibinfo {pages} {155455} (\bibinfo {year}
  {2012}{\natexlab{b}})}\BibitemShut {NoStop}%
\bibitem [{\citenamefont {Zar\'and}\ and\ \citenamefont
  {Udvardi}(1996)}]{Zarand96}%
  \BibitemOpen
  \bibfield  {author} {\bibinfo {author} {\bibfnamefont {G.}~\bibnamefont
  {Zar\'and}}\ and\ \bibinfo {author} {\bibfnamefont {L.}~\bibnamefont
  {Udvardi}},\ }\href {\doibase 10.1103/PhysRevB.54.7606} {\bibfield  {journal}
  {\bibinfo  {journal} {Phys. Rev. B}\ }\textbf {\bibinfo {volume} {54}},\
  \bibinfo {pages} {7606} (\bibinfo {year} {1996})}\BibitemShut {NoStop}%
\bibitem [{\citenamefont {Kettemann}(2004)}]{Kettemann04}%
  \BibitemOpen
  \bibfield  {author} {\bibinfo {author} {\bibfnamefont {S.}~\bibnamefont
  {Kettemann}},\ }in\ \href@noop {} {\emph {\bibinfo {booktitle} {Quantum
  Information and Decoherence in Nanosystems}}},\ \bibinfo {editor} {edited by\
  \bibinfo {editor} {\bibfnamefont {D.~C.}\ \bibnamefont {Glattli}}, \bibinfo
  {editor} {\bibfnamefont {M.}~\bibnamefont {Sanquer}}, \ and\ \bibinfo
  {editor} {\bibfnamefont {J.~T.~T.}\ \bibnamefont {Van}}}\ (\bibinfo
  {publisher} {The Gioi Publishers},\ \bibinfo {address} {Hannoi},\ \bibinfo
  {year} {2004})\ p.\ \bibinfo {pages} {259},\ \bibinfo {note}
  {(cond-mat/0409317)}\BibitemShut {NoStop}%
\bibitem [{\citenamefont {Kettemann}\ and\ \citenamefont
  {Mucciolo}(2006)}]{Kettemann06}%
  \BibitemOpen
  \bibfield  {author} {\bibinfo {author} {\bibfnamefont {S.}~\bibnamefont
  {Kettemann}}\ and\ \bibinfo {author} {\bibfnamefont {E.~R.}\ \bibnamefont
  {Mucciolo}},\ }\href@noop {} {\bibfield  {journal} {\bibinfo  {journal}
  {Pis'ma v ZhETF}\ }\textbf {\bibinfo {volume} {83}},\ \bibinfo {pages} {284}
  (\bibinfo {year} {2006})},\ \bibinfo {note} {[JETP Letters {\bf 83}, 240
  (2006); cond-mat:0509251]}\BibitemShut {NoStop}%
\bibitem [{\citenamefont {Kettemann}\ and\ \citenamefont
  {Mucciolo}(2007)}]{Kettemann07}%
  \BibitemOpen
  \bibfield  {author} {\bibinfo {author} {\bibfnamefont {S.}~\bibnamefont
  {Kettemann}}\ and\ \bibinfo {author} {\bibfnamefont {E.~R.}\ \bibnamefont
  {Mucciolo}},\ }\href@noop {} {\bibfield  {journal} {\bibinfo  {journal}
  {Phys. Rev. B}\ }\textbf {\bibinfo {volume} {75}},\ \bibinfo {eid} {184407}
  (\bibinfo {year} {2007})}\BibitemShut {NoStop}%
\bibitem [{\citenamefont {Lee}\ \emph {et~al.}(2006)\citenamefont {Lee},
  \citenamefont {Nagaosa},\ and\ \citenamefont {Wen}}]{MeanFieldForSpinReview}%
  \BibitemOpen
  \bibfield  {author} {\bibinfo {author} {\bibfnamefont {P.~A.}\ \bibnamefont
  {Lee}}, \bibinfo {author} {\bibfnamefont {N.}~\bibnamefont {Nagaosa}}, \ and\
  \bibinfo {author} {\bibfnamefont {X.}~\bibnamefont {Wen}},\ }\href@noop {}
  {\bibfield  {journal} {\bibinfo  {journal} {Rev. Mod. Phys.}\ }\textbf
  {\bibinfo {volume} {78}},\ \bibinfo {pages} {17} (\bibinfo {year}
  {2006})}\BibitemShut {NoStop}%
\bibitem [{\citenamefont {Aleiner}\ \emph {et~al.}(2002)\citenamefont
  {Aleiner}, \citenamefont {Brouwer},\ and\ \citenamefont
  {Glazman}}]{Aleiner02PhysRep}%
  \BibitemOpen
  \bibfield  {author} {\bibinfo {author} {\bibfnamefont {I.~L.}\ \bibnamefont
  {Aleiner}}, \bibinfo {author} {\bibfnamefont {P.~W.}\ \bibnamefont
  {Brouwer}}, \ and\ \bibinfo {author} {\bibfnamefont {L.~I.}\ \bibnamefont
  {Glazman}},\ }\href@noop {} {\bibfield  {journal} {\bibinfo  {journal} {Phys.
  Rep.}\ }\textbf {\bibinfo {volume} {358}},\ \bibinfo {pages} {309} (\bibinfo
  {year} {2002})}\BibitemShut {NoStop}%
\bibitem [{\citenamefont {Ullmo}(2008)}]{UllmoRPP2008}%
  \BibitemOpen
  \bibfield  {author} {\bibinfo {author} {\bibfnamefont {D.}~\bibnamefont
  {Ullmo}},\ }\href@noop {} {\bibfield  {journal} {\bibinfo  {journal} {Rep.
  Prog. Phys.}\ }\textbf {\bibinfo {volume} {71}},\ \bibinfo {pages} {026001}
  (\bibinfo {year} {2008})}\BibitemShut {NoStop}%
\bibitem [{\citenamefont {Hewson}(1993{\natexlab{b}})}]{HewsonBook2}%
  \BibitemOpen
  \bibfield  {author} {\bibinfo {author} {\bibfnamefont {A.}~\bibnamefont
  {Hewson}},\ }\href@noop {} {\emph {\bibinfo {title} {The Kondo Problem to
  Heavy Fermions}}}\ (\bibinfo  {publisher} {Cambridge University Press},\
  \bibinfo {address} {Cambridge},\ \bibinfo {year} {1993})\ pp.\ \bibinfo
  {pages} {113, 413--417}\BibitemShut {NoStop}%
\end{thebibliography}

%merlin.mbs apsrev4-1.bst 2010-07-25 4.21a (PWD, AO, DPC) hacked
%Control: key (0)
%Control: author (8) initials jnrlst
%Control: editor formatted (1) identically to author
%Control: production of article title (-1) disabled
%Control: page (0) single
%Control: year (1) truncated
%Control: production of eprint (0) enabled

%

\end{document}